\documentclass[iop]{emulateapj}

\usepackage{natbib}
\usepackage{amsmath,amssymb,mathrsfs}
\usepackage{longtable}
\usepackage{here}

\shortauthors{HAYASHI \& CHIBA.}
\shorttitle{The prolate dark halo of M31}

\begin{document}

\title{The prolate dark matter halo of the Andromeda galaxy}

\author{Kohei~Hayashi\altaffilmark{1} and
	Masashi~Chiba\altaffilmark{1}}

\altaffiltext{1}{Astronomical Institute, Tohoku University,
Aoba-ku, Sendai 980-8578, Japan \\E-mail: {\it k.hayasi@astr.tohoku.ac.jp; chiba@astr.tohoku.ac.jp}}

\begin{abstract}
We present new limits on the global shape of the dark matter halo in the Andromeda galaxy using and generalizing non-spherical mass models developed by Hayashi \& Chiba and compare our results with theoretical predictions of cold dark matter (CDM) models.
This is motivated by the fact that CDM models predict non-spherical virialized dark halos, which reflect the process of mass assembly in the galactic scale.
Applying our models to the latest kinematic data of globular clusters and dwarf spheroidal galaxies in the Andromeda halo, we find that the most plausible cases for Andromeda yield a prolate shape for its dark halo, irrespective of assumed density profiles.
We also find that this prolate dark halo in Andromeda is consistent with theoretical predictions in which the satellites are distributed anisotropically and preferentially located along major axes of their host halos. It is a reflection of the intimate connection between galactic dark matter halos and the cosmic web. Therefore, our result is profound in understanding internal dynamics of halo tracers in Andromeda, such as orbital evolutions of tidal stellar streams, which play important roles in extracting the abundance of CDM subhalos through their dynamical effects on stream structures.
\end{abstract}


\section{INTRODUCTION}
The Milky Way (MW) and its nearest neighbor Andromeda (M31) provide a unique laboratory to test cold dark matter (CDM) theory of galaxy formation and evolution. 
In particular, CDM models as a current paradigm of structure formation in the Universe predict a universal density distribution for a galaxy-sized halo as well as the presence of numerous subgalactic halos in it, as a consequence of hierarchical assembly process of dark matter. Thus, it is of importance to derive how dark matter is actually distributed in a galaxy scale such as that of the MW and M31 to get useful insight into the role of dark halos in galactic structure and evolution in the framework of CDM models.

High-resolution, CDM-based $N$-body simulations have been successful in predicting the large scale structure of galaxy distribution on spatial scales greater than a few megaparsecs, whereas the models are yet unsuccessful in fully explaining smaller spatial scales, i.e., galaxy scales such as the MW, M31, and their substructures. 
For example, the ``missing satellite problem'' is one of the open questions in CDM theory: the large discrepancy between the small number of observed satellites around a galaxy such as the MW and M31 and the large number of predicted surviving CDM subhalos (Moore et al. 1999; Klypin et al. 1999; Diemand et al. 2008; Springel et al. 2008). 
In recent years, this discrepancy is thought to be due to the astrophysical processes; UV reionization, baryonic cooling, supernovae, and gas accretion affect star formation in dark matter halos (e.g., Bullock et al. 2000; Benson et al. 2002; Kaufmann et al. 2008). 
Furthermore, hydrodynamical simulations which include gas dynamics and radiative transfer have been able to reproduce some of the observed luminosity and metallicity properties of galaxies (e.g., Springel 2010). 
Although the complete understanding of star formation process that reproduce the observed properties of satellite galaxies remains uncertain, the solution to the missing satellite problem will provide important implications for baryonic and/or non-baryonic physics in galaxy formation and evolution.

In addition, the spatial distribution of satellite galaxies relative to their host galaxies may hold important clues to the formation history of galaxies. 
This is especially true for CDM models in which the dark matter halos of galaxies are mildly flattened, and galaxy formation and mass accretion occur within filaments. 
Thus, the key to understanding the origin of the anisotropic satellites distribution lies in the connection between halos and the cosmic web and, in particular, in the way in which satellites are accreted into the main halo. 

Some of early studies on the location of satellite galaxies suggested that satellites pertaining to host disk galaxies with a projected radius of $r_{p}\leq50$~kpc are located preferentially near the minor axes of their hosts (e.g., Holmberg 1969; Zaritsky et al. 1997). 
On the other hand, several subsequent studies found that satellites tend to be aligned with major axes of their hosts (Valtonen et al. 1978) or concluded little evidence for such a preferential alignment (e.g., Hawley \& Peebles 1975; Sharp et al. 1979). 
Recently, the statics for the spatial distribution of satellites have been greatly improved with a large number of available sample galaxies provided by several observational surveys. 
Agustsson \& Brainerd (2010) investigated the location of the satellites with respect to their host spheroidal galaxies at redshift $0.01<z<0.15$ in the Sloan Digital Sky Survey Data Release 7 (Abazajian et al. 2009). 
They found that within $r_{p}\leq500$~kpc, the degree of anisotropic spatial distribution was greatest for the satellites of red, high-mass hosts with low star formation rates, while the location of the satellites of blue, low-mass hosts with high star formation rates were consistent with isotropic distribution.
Nierenberg~et~al. (2011) have used high-resolution {\it Hubble Space Telescope} imaging from the GOODS fields and have studied the spatial distribution of faint satellites at moderate redshift ($0.1<z<0.8$). They found that the satellites are preferentially aligned along major axes of their hosts.
This issue has also been investigated for the MW and M31.
Many previous studies reported that the spatial distribution of satellites in the MW and M31 is manifestly anisotropic with the majority of the satellites found in a flattened structure nearly perpendicular to the disk component (e.g., Lynden-Bell 1982; Hartwick 2000; Majewski 1994; Kroupa et al. 2005; Koch \& Grebel 2006; Ibata et al. 2013; Conn et al. 2013). 
In particular, Ibata et al. (2013) discovered the existence of a vast planer subgroup of satellites in M31 and also the coherent rotational motion of satellites within this structure.

Several numerical simulations have tested if this anisotropic satellite distribution can be understood within the framework of CDM cosmology (Kang et al. 2005; Zentner et al. 2005; Libeskind et al. 2005; Deason et al. 2011).
According to Libeskind et al. (2005), who have used high-resolution $N$-body simulation and a semi-analytic model to follow the formation of satellite galaxies, brightest satellites are distributed along the elongated disk-like structure, even though the distribution of most massive subhalos is almost spherical. 
Interestingly, this characteristic structure has its long axis aligned with the major axis of the host dark matter halo as a consequence of the preferential infall of satellites along a few filaments of the cosmic web. 

Accordingly, it is crucial to derive and discuss the detailed structure of dark halos in galaxies in comparison with theoretical predictions. 
In this work, we investigate the global shape and density profile of the dark halo in M31. 
We note that most of the existing mass models for the M31 dark halo have assumed spherical symmetry for the purpose of simply estimating its total mass (Evans \& Wilkinson 2000; Watkins et al. 2010; Veljanoski et al. 2013). 
On the other hand, CDM models predict non-spherical virialized dark halos in this galaxy scale. Thus, to obtain the more realistic mass distribution of the dark halo in M31 and compare our results with theoretical prediction of CDM models, we adopt and generalize the axisymmetric mass models developed by Hayashi \& Chiba (2012, hereafter HC12) and apply these generalized models to the latest kinematic data of globular clusters (GCs) and dwarf spheroidal galaxies (dSphs) in the halo of M31.

This paper is organized as follows. In Section 2, we briefly describe the data set we use as the halo tracers of M31 and models for density profiles of stellar and dark halo components based on Jeans analysis. The method of a maximum likelihood analysis to be applied to the observed halo tracers in M31 is also presented.   
In Section 3, we present the results of maximum likelihood analysis.
Finally, in Section 4, we discuss our results and their implications for CDM theory and present our conclusions.
 \begin{figure}[!t]
\begin{center}
  \includegraphics[scale=0.7]{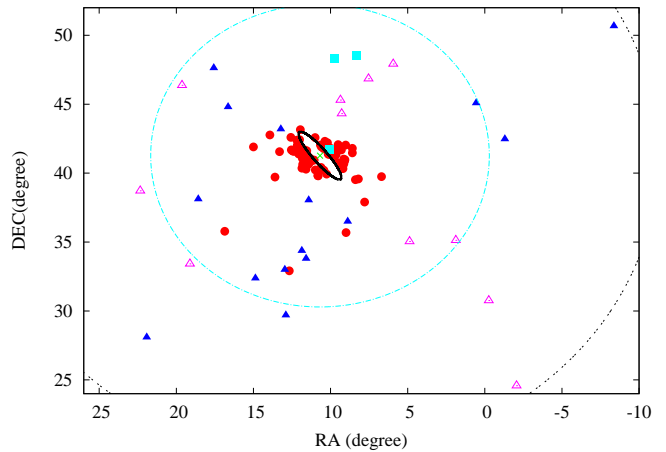}
 \end{center}
 \caption{Spatial distribution of globular clusters (GCs) and dwarf spheroidal (dSph) satellites in M31. The M31 center is indicated by green cross. A thick black ellipse is the M31 disk/halo boundary as defined by Racine (1991). The cyan dot-dashed and black two-dotted circles centered on M31 denote the projected radii of 150 and 300 kpc, respectively. Red dots and blue filled triangles show the GC and dSph samples here, respectively, whereas the empty triangles and cyan filled squares denote the dSphs and dEs, respectively, which we do not adopt for our analysis.}
 \label{fig:fig1}
\end{figure}

\section{THE DATA AND THE MODEL}
\subsection{The Observational Data Sets}
In this subsection, we present the observational data for GCs and dSphs in M31 which we use as the halo tracers. 
We use a sample of 91 GCs and 15 dSphs in M31. 
For the sample of GCs, we adopt the Revised Bologna Catalogue of M31 GCs and candidates (RBC v.4\footnote{Available from http://www.bo.astro.it/M31}; Galleti et al. 2004, 2006, 2009), which is the latest and most comprehensive M31 GC catalog so far. 
For the sample of dSphs, we use published data in Tollerud~et~al.~(2012; 15 target objects: And I, III, V, VII, IX, X, XI, XII, XIII, XIV, XV, XVI, XVIII, XXI and XXII). 
We take the position of the center of M31 to be ${\rm (R.A., decl.)} = (10^{\circ}.7, 41^{\circ}.2)$ at a distance of 785 kpc, its line-of-sight velocity to be $-123$ km~s$^{-1}$ in the Galactic rest frame, and its pitch and inclination angles to be P.A. $=38^{\circ}$ and $i=12^{\circ}.5$, respectively (see, e.g., Durrell et al. 2004; McConnachie et al. 2005; McConnachie \& Irwin 2006; de Vaucouleurs 1958; Rubin \& D'Odorico 1969). 
Tables 1 and 2 show the referred data set for GCs and dSphs in M31: object number, distance from M31's center on the $(x,y)$ coordinate, uncorrected line-of-sight velocity, and equatorial coordinates (R.A., decl.). 
Figure 1 shows their spatial distribution. 
Red dots and blue filled triangles denote the adopted GCs and dSphs, respectively.  A thick black ellipse is the M31 disk/halo boundary, which has a long axis of 30~kpc and a short axis of 6~kpc, as defined by Racine (1991). Using this boundary, we adopt 91 halo component GCs in our sample. 
Since we assume axisymmetry in this work, we analyze the velocity data by folding up the stellar distribution into the first quadrant for these objects in M31's halo.

In converting the heliocentric velocities given in Tables 1 and 2 to the Galactocentric ones, we assume a circular velocity ($V_{\rm lsr}$) of 220 km~s$^{-1}$ at the Galactocentric radius of the Sun ($R_{\odot}$=8.0kpc). 
The values of the peculiar motion of the Sun $(U,V,W) = (11.4,12.24,7.25)$ km~s$^{-1}$, where $U$ is directed inward to the Galactic center, $V$ is positive in the direction of Galactic rotation at the position of the Sun, and $W$ is positive towards the North Galactic Pole, are adopted from Sch\"{o}nrich et al. (2010). 
Hereafter, $v_{\rm los}$ is the radial velocity in the Galactic standard of rest (hereafter GSR; i.e., the radial velocity component along the star--Sun direction, corrected for Galactic rotation). 
\begin{longtable}{lccccc}[t!!]
\label{tb:tab1}
\tablecolumns{6}
\tablewidth{4.0in}
\tablecaption{The List of the M31 Globular Clusters Used in This Work.}
\tablehead{Object No.  & $x$   & $y$  &$v_{\rm los}$ & R.A. & Decl. \\
                                       &  (kpc) & (kpc)  & (km~s$^{-1}$) & (deg) & (deg) }
1 & -43.59 & -22.39 & -147.0  $ \pm20.0 $ & 13.93 & 42.77 \\ 
2 & -43.20 & -41.23 & -272.0  $ \pm54.0 $ & 15.00 & 41.90 \\ 
3 & -31.09 &   2.15 &  33.0  $ \pm30.0 $ & 11.96 & 43.15 \\ 
4 & -30.26 &  -9.23 & -295.0  $ \pm26.0 $ & 12.58 & 42.60 \\ 
5 & -25.25 & -25.92 & -154.0  $ \pm30.0 $ & 13.31 & 41.56 \\ 
6 & -24.83 &  -6.76 & -97.0 $ \pm12.0 $ & 12.19 & 42.39 \\ 
7 & -23.63 &  -4.51 & -258.0 $ \pm11.0 $ & 12.01 & 42.43 \\ 
8 & -20.53 &  -9.44 & -488.0 $ \pm26.0 $ & 12.15 & 42.03 \\ 
9 & -20.13 & -16.55 & -338.0 $ \pm17.0 $ & 12.54 & 41.68 \\ 
10 & -19.63 & -10.44 & -49.0 $ \pm22.0 $ & 12.17 & 41.93 \\ 
11 & -18.72 & -16.36 &  21.0 $ \pm15.0 $ & 12.47 & 41.61 \\ 
12 & -18.08 & -15.97 & -162.0 $ \pm16.0 $ & 12.42 & 41.59 \\ 
13 & -17.24 & -15.02 & -358.0 $ \pm48.0 $ & 12.32 & 41.59 \\ 
14 & -16.88 &  -9.52 & -227.0 $ \pm5.0 $ & 11.99 & 41.81 \\ 
15 & -16.84 &   7.61 &  16.0 $ \pm26.0 $ & 11.00 & 42.58 \\ 
16 & -15.81 & -11.13 & -333.0 $ \pm23.0 $ & 12.04 & 41.68 \\ 
17 & -14.55 & -11.47 & -437.0 $ \pm20.0 $ & 12.00 & 41.59 \\ 
18 & -13.98 & -13.01 & -206.0 $ \pm41.0 $ & 12.06 & 41.49 \\ 
19 & -13.39 &  -7.39 & -21.0 $ \pm58.0 $ & 11.71 & 41.71 \\ 
20 & -12.99 & -12.41 & -16.0 $ \pm59.0 $ & 11.98 & 41.46 \\ 
21 & -12.98 & -13.88 & -252.0 $ \pm46.0 $ & 12.07 & 41.39 \\ 
22 & -12.12 &  -8.92 & -325.0 $ \pm12.0 $ & 11.74 & 41.57 \\ 
23 & -11.11 &  -6.24 & -302.0 $ \pm12.0 $ & 11.54 & 41.63 \\ 
24 & -10.46 & -10.44 & -331.0 $ \pm10.0 $ & 11.76 & 41.40 \\ 
25 &  -9.23 & -13.29 & -228.0 $ \pm50.0 $ & 11.86 & 41.20 \\ 
26 &  -8.60 &  12.16 & -63.0 $ \pm15.0 $ & 10.37 & 42.31 \\ 
27 &  -8.07 &   6.75 & -164.0 $ \pm50.0 $ & 10.66 & 42.04 \\ 
28 &  -7.98 &  -8.42 & -86.0 $ \pm2.0 $ & 11.53 & 41.35 \\ 
29 &  -7.93 &  -8.84 & -253.0 $ \pm9.0 $ & 11.55 & 41.33 \\ 
30 &  -7.84 & -44.59 & -381.0 $ \pm15.0 $ & 13.60 & 39.72 \\ 
31 &  -7.24 & -14.08 &  12.0 $ \pm55.0 $ & 11.82 & 41.05 \\ 
32 &  -6.14 &   6.35 & -251.0 $ \pm20.0 $ & 10.60 & 41.91 \\ 
33 &  -6.01 &  13.09 &  50.0 $ \pm12.0 $ & 10.20 & 42.20 \\ 
34 &  -5.00 &   6.90 & -252.0 $ \pm13.0 $ & 10.51 & 41.87 \\ 
35 &  -4.55 &   7.69 & -385.0 $ \pm56.0 $ & 10.45 & 41.88 \\ 
36 &  -3.92 & -11.73 & -152.0 $ \pm45.0 $ & 11.54 & 40.97 \\ 
37 &  -3.89 &  -8.61 & -56.0 $ \pm5.0 $ & 11.35 & 41.11 \\ 
38 &  -3.35 &   6.95 & -98.0 $ \pm49.0 $ & 10.44 & 41.77 \\ 
39 &  -2.66 & -17.48 & -397.0 $ \pm38.0 $ & 11.81 & 40.64 \\ 
40 &  -2.31 & -14.80 & -297.0 $ \pm78.0 $ & 11.64 & 40.74 \\ 
41 &  -2.09 &   8.33 & -310.0 $ \pm34.0 $ & 10.30 & 41.76 \\ 
42 &  -0.78 & -13.48 & -373.0 $ \pm32.0 $ & 11.50 & 40.71 \\ 
43 &  -0.71 &   8.46 & -351.0 $ \pm1.0 $ & 10.23 & 41.69 \\ 
44 &  -0.49 &  -7.65 & -597.0 $ \pm5.0 $ & 11.15 & 40.95 \\ 
45 &  -0.46 & -10.11 & -411.0 $ \pm31.0 $ & 11.29 & 40.84 \\ 
46 &  -0.07 & -20.29 & -561.0 $ \pm30.0 $ & 11.85 & 40.36 \\ 
47 &   0.19 &   9.86 & -60.0 $ \pm24.0 $ & 10.11 & 41.70 \\ 
48 &   0.25 &  12.05 & -143.0 $ \pm23.0 $ &  9.98 & 41.80 \\ 
49 &   0.30 &  -6.80 & -512.0 $ \pm9.0 $ & 11.06 & 40.95 \\ 
50 &   0.33 & -13.04 & -176.0 $ \pm28.0 $ & 11.42 & 40.66 \\ 
51 &   0.39 & -13.63 & -157.0 $ \pm21.0 $ & 11.45 & 40.63 \\ 
52 &   0.49 &   9.21 & -178.0 $ \pm52.0 $ & 10.13 & 41.65 \\ 
53 &   0.55 &   9.69 & -249.0 $ \pm22.0 $ & 10.10 & 41.67 \\ 
54 &   0.62 &   9.94 & -299.0 $ \pm35.0 $ & 10.09 & 41.68 \\ 
55 &   0.96 &   8.93 & -356.0 $ \pm52.0 $ & 10.13 & 41.62 \\ 
56 &   1.32 &  19.47 & -27.0 $ \pm33.0 $ &  9.51 & 42.07 \\ 
57 &   2.76 &   6.96 & -409.0 $ \pm12.0 $ & 10.16 & 41.42 \\ 
58 &   2.81 &   7.79 & -237.0 $ \pm1.0 $ & 10.11 & 41.46 \\ 
59 &   3.00 & -18.10 & -364.0 $ \pm1.0 $ & 11.59 & 40.28 \\ 
60 &   3.54 &  13.54 & -405.0 $ \pm43.0 $ &  9.75 & 41.67 \\ 
61 &   3.63 &   6.72 & -363.0 $ \pm3.0 $ & 10.14 & 41.36 \\ 
62 &   3.97 &   7.50 & -369.0 $ \pm12.0 $ & 10.07 & 41.38 \\ 
63 &   5.65 &   9.97 & -206.0 $ \pm18.0 $ &  9.86 & 41.39 \\ 
64 &   5.76 &  24.43 & -215.0 $ \pm30.0 $ &  9.02 & 42.04 \\ 
65 &   7.20 &  18.54 & -316.0 $ \pm35.0 $ &  9.29 & 41.69 \\ 
66 &   7.42 & -112.28 & -291.0 $ \pm2.0 $ & 16.86 & 35.78 \\ 
67 &   7.97 &  11.95 & -371.0 $ \pm23.0 $ &  9.64 & 41.35 \\ 
68 &   8.62 & -12.01 & -330.0 $ \pm26.0 $ & 10.99 & 40.23 \\ 
69 &   9.01 &   9.43 & -341.0 $ \pm13.0 $ &  9.74 & 41.17 \\ 
70 &   9.79 &  -6.50 & -467.0 $ \pm13.0 $ & 10.62 & 40.41 \\ 
71 &  10.61 &  -8.04 & -348.0 $ \pm26.0 $ & 10.67 & 40.30 \\ 
72 &  10.72 & -10.47 & -285.0 $ \pm18.0 $ & 10.81 & 40.18 \\ 
73 &  11.60 &  -6.87 & -293.0 $ \pm26.0 $ & 10.56 & 40.29 \\ 
74 &  11.99 &  27.10 & -181.0 $ \pm30.0 $ &  8.59 & 41.80 \\ 
75 &  13.63 &  -6.21 & -360.0 $ \pm2.0 $ & 10.43 & 40.21 \\ 
76 &  14.46 & -13.73 & -315.0 $ \pm2.0 $ & 10.82 & 39.82 \\ 
77 &  14.88 & -13.67 & -278.0 $ \pm22.0 $ & 10.80 & 39.80 \\ 
78 &  15.50 &  24.35 & -381.0 $ \pm26.0 $ &  8.59 & 41.47 \\ 
79 &  15.68 &  14.50 & -403.0 $ \pm93.0 $ &  9.15 & 41.02 \\ 
80 &  15.78 &   8.25 & -539.0 $ \pm12.0 $ &  9.50 & 40.73 \\
81 &  16.81 &  14.95 & -307.0 $ \pm18.0 $ &  9.07 & 40.97 \\ 
82 &  17.53 &  14.08 & -424.0 $ \pm23.0 $ &  9.09 & 40.89 \\ 
83 &  18.08 &  -6.62 & -188.0 $ \pm26.0 $ & 10.25 & 39.93 \\ 
84 &  19.58 &  11.56 & -240.0 $ \pm47.0 $ &  9.14 & 40.66 \\ 
85 &  22.72 &   8.15 & -408.0 $ \pm11.0 $ &  9.19 & 40.33 \\ 
86 &  38.19 &  10.01 & -313.0 $ \pm17.0 $ &  8.39 & 39.52 \\ 
87 &  39.24 &  12.61 & -332.0 $ \pm3.0 $ &  8.19 & 39.58 \\ 
88 &  50.02 &  30.16 & -219.0 $ \pm15.0 $ &  6.70 & 39.75 \\ 
89 &  60.68 &   2.77 & -458.0 $ \pm23.0 $ &  7.79 & 37.90 \\ 
90 &  73.45 & -91.53 & -355.0 $ \pm2.0 $ & 12.68 & 32.92 \\ 
91 &  74.34 & -28.99 & -358.0 $ \pm2.0 $ &  9.00 & 35.68 \\ 
\end{longtable}

\subsection{Axisymmetric Jeans Equations}
A simple but more realistic representation of M31 than previous spherically symmetric models is an axisymmetric mass distribution. 
HC12 assumed in their axisymmetric mass models that the phase-space distribution function of the halo tracers is given in  the form $f(E,L_z)$, where $E$ and $L_z$ denote binding energy and angular momentum component toward the symmetry axis, respectively. Thus, for velocity dispersion $(\overline{v^2_R}, \overline{v^2_{\phi}}, \overline{v^2_z})$ of the halo tracers in cylindrical coordinates $(R,\phi,z)$, this assumption corresponds to $\overline{v^2_R}=\overline{v^2_z}$.
In this paper, we relax this constraint and consider a velocity anisotropy parameter, $\beta_z=1-\overline{v^2_z}/\overline{v^2_R}$, where we assume the constant anisotropy,  $\beta_z=$ constant. In this case, the Jeans equations are written as
\begin{equation}
\overline{v^2_z} =  \frac{1}{\nu(R,z)}\int^{\infty}_z \nu\frac{\partial \Phi}{\partial z}dz,
\label{eq:eq1}
\end{equation}
\begin{equation}
\overline{v^2_{\phi}} = \frac{1}{1-\beta_z} \Biggl[ \overline{v^2_z} + \frac{R}{\nu}\frac{\partial(\nu\overline{v^2_z})}{\partial R} \Biggr] + 
R \frac{\partial \Phi}{\partial R},
\label{eq:eq2}
\end{equation}
where $\nu$ is the density of stellar population as halo tracers and $\Phi$ are a gravitational potential which is represented by three components: a bulge, disk and dark halo.
We assume a Hernquist model for the bulge with a total mass of $3.3\times10^{10} M_{\odot}$ and a scale length of $0.61$~kpc, the disk having an exponential surface-density distribution with a total mass of $2.0\times 10^{11} M_{\odot}$ and a scale length of $5.4$~kpc (Geehan et al. 2006), and the dark halo having a power-law density distribution, which will be described in more detail in Section~2.4.
One of the benefits in this axisymmetric model is that the change of velocity anisotropy between $\overline{v^2_z}$ (or $\overline{v^2_R}$) and $\overline{v^2_{\phi}}$ with varying radii is fully taken into account.
In order to compare these solutions with observed stellar kinematics in M31, we derive the line-of-sight velocity dispersion from $\overline{v^2_R}$, $\overline{v^2_{\phi}}$ and $\overline{v^2_z}$, taking into account inclination between the line of sight and the galactic plane, following the method given in Tempel \& Tenjes (2006). 
These velocity dispersions are provided by the second moments that separate into the contribution of ordered and random motions, as defined by $\overline{v^2} = \sigma^2 + \overline{v}^2$.
\begin{deluxetable}{lccccc}[H]
\label{tb:tab2}
\tablecolumns{6}
\tablewidth{3.2in}
\tablecaption{The List of the M31 Dwarf Galaxies Used in This Work\tablenotemark{*}}
\tablehead{Object No.  & $x$   & $y$  &$v_{\rm los}$ & R.A. & Decl. \\
                                       &  (kpc) & (kpc)  & (km~s$^{-1}$) & (deg) & (deg) }
\startdata
1 &  28.69 & -35.10 & -376.3 $ \pm10.2 $ & 11.42 & 38.04 \\ 
2 &  66.51 & -20.95 & -344.3 $ \pm9.3 $ &  8.89 & 36.50 \\ 
3 & -125.84 & -20.25 & -397.3 $ \pm10.5 $ & 17.57 & 47.63 \\ 
4 &  56.94 & 276.23 & -307.2 $ \pm13.0 $ & -8.37 & 50.68 \\ 
5 & -42.16 & -11.12 & -209.4 $ \pm10.9 $ & 13.22 & 43.20 \\ 
6 & -88.12 & -34.35 & -164.1 $ \pm6.4 $ & 16.64 & 44.80 \\ 
7 &  73.00 & -72.51 & -461.8 $ \pm3.7 $ & 11.58 & 33.80 \\ 
8 &  64.50 & -70.69 & -525.3 $ \pm3.4 $ & 11.86 & 34.37 \\ 
9 &  70.12 & -93.83 & -185.4 $ \pm5.8 $ & 12.96 & 33.00 \\ 
10 & 106.34 & -120.52 & -480.6 $ \pm5.3 $ & 12.90 & 29.70 \\ 
11 & -32.44 & -111.30 & -323.0 $ \pm4.0 $ & 18.58 & 38.12 \\ 
12 &  61.03 & -119.23 & -367.3 $ \pm3.8 $ & 14.87 & 32.38 \\ 
13 &  43.95 & 140.52 & -332.1 $ \pm9.7 $ &  0.56 & 45.09 \\ 
14 &  87.92 & 138.78 & -361.4 $ \pm7.2 $ & -1.30 & 42.47 \\ 
15 &  50.00 & -225.99 & -126.8 $ \pm3.5 $ & 21.92 & 28.09
\enddata
\tablenotetext{*}{Taken from Tollerud et al. (2012)}
\end{deluxetable}
 
\subsection{Density Profile of the Halo Tracers}
The Jeans equations allow us to obtain a unique solution for the dark halo mass profile if we know $\overline{v^2_z}$ (or $\overline{v^2_R}$), $\overline{v^2_{\phi}} $, and $\nu$ of the halo tracers, although this solution is not guaranteed to produce a positive phase space distribution function. 
To obtain the volume density $\nu$ of GCs and dSphs in M31, we assume that their surface number density falls off like a power law,
$\Sigma_{\ast}(m^{\prime}_{\ast})\propto m^{\prime\gamma}_{\ast}$, where $m^{\prime2}_{\ast}= x^2+y^2/q^{\prime2}$, $q^{\prime}$ is the projected axial ratio and $(x,y)$ are the coordinates along and perpendicular to the M31 disk component, respectively. 
We obtain a power-law index, $\gamma$, by fitting to the surface number density through the use of the $\chi^2$ method. 
Figure~2 shows the spherically averaged surface number density, where the solid line is the  best-fit model result with $\gamma=-3.00\pm0.05$.
The three-dimensional number density, $\nu$, is simply derived by deprojection of the surface density profile, i.e., $\nu(m_{\ast})\propto m_{\ast}^{\gamma-1}$, where $m_{\ast}^2=R^2+z^2/q^2$ and $q$ is the intrinsic axial ratio. 
Projected axial ratio $q^{\prime}$ is related to $q$ through inclination angle $i$ as  $q^{\prime 2}=\cos^2i + q^2\sin^2i$, where $i=90^{\circ}$ when a galaxy is edge-on and $i=0^{\circ}$ for face-on. 
 \begin{figure}[t!]
\begin{center}
  \includegraphics[scale=0.7]{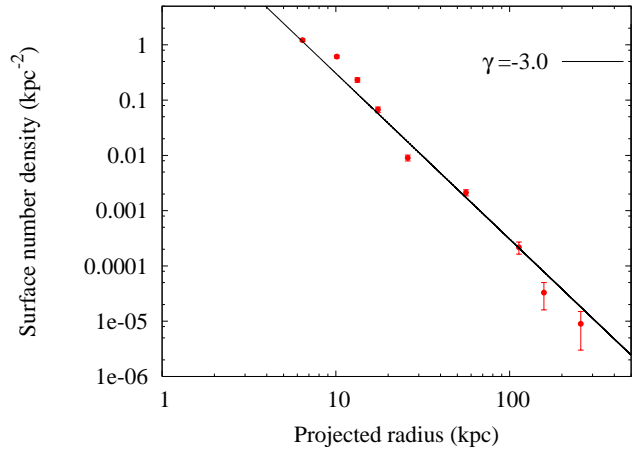}
 \end{center}
 \caption{Surface number density of halo tracers in M31 as a function of projected radius. Filled circles with error bar denote the observed number density of the GC and dSph tracers. The best-fit power-law profile is plotted as a solid line. }
 \label{fig:fig2}
\end{figure}

To obtain the projected axial ratio of the stellar component, $q^{\prime}$, we apply the ``reduced'' inertia tensor method as implemented by Allgood et al. (2006) in their numerical simulation. This tensor is defined as
\begin{equation}
I_{ij} = \sum_{n}\frac{x_{i,n}x_{j,n}}{r^2_n},
\end{equation}
where 
\begin{equation}
r_n = \sqrt{x^2_n + y^2_n/q^{\prime2}} 
\end{equation}
is the elliptical distance in the eigenvector coordinate system from the center to the $n$th halo tracers. In this case, the eigenvalues ($\lambda_a \leq \lambda_b$) determine the projected axial ratio, that is, $q^{\prime}=\sqrt{\lambda_b}/\sqrt{\lambda_a}$. The method used here begins by determining $I$ with $q^{\prime}=1$, including all tracers within some radius. Subsequently, a new value for $q^{\prime}$ is determined and the area of analysis is deformed along the eigenvectors in proportion to the eigenvalues. After the deformation of the original spherical region, $I$ is calculated once again, but now using the newly determined $q^{\prime}$ and only including the tracers found in the new ellipsoidal region. The iterative process is repeated until convergence is achieved. Convergence is achieved when the variance in axial ratio is less than a given tolerance (see Section~3 in Allgood et al. 2006).
Applying this method to GC and dSph samples in M31, the axial ratio in their density distribution yields $q^{\prime}\simeq1.18\pm0.20$, the random error of which is estimated by the Monte Carlo method.
However, because of small amount of available photometric and kinematic data in the M31 halo as shown in Figure 1, the results of both $\gamma$ and $q^{\prime}$ are still fairly uncertain. 

\subsection{Dark Matter Halo Model}
For the dark matter halo, we assume the following power-law form:
 \begin{equation}
 \rho(R,z) = \rho_0 \Bigl(\frac{m}{b_{\rm halo}} \Bigr)^{\alpha}\Bigl[1+\Bigl(\frac{m}{b_{\rm halo}} \Bigr)^2 \Bigr]^{\delta},
 \label{eq:eq2}
 \end{equation}
\begin{equation}
m^2=R^2+z^2/Q^2,
\label{eq:eq3}
 \end{equation}
where $\rho_0$ is a density normalization, $b_{\rm halo}$ is a scale length in the spatial distribution, and $Q$ is an axial ratio of the dark halo.
\begin{deluxetable}{ccccc}[H]
\label{tb:tab3}
\tablecolumns{5}
\tablewidth{3.0in}
\tablecaption{Maximum Likelihood Results for the M31 Halo}
\tablehead{ Halo Model & $Q$ & $b_{\rm halo}$ & $M_{\rm \leq200kpc}$ & $\beta_z$ \\
                                       &          &       (kpc)             & ($10^{12}M_{\odot}$)   &
}
\startdata
           SIS      &  $ 1.60^{+0.38}_{-0.51}$ & $ >8.1 $     & $ 1.39^{+0.42}_{-0.34}$  &  $-0.46^{+0.28}_{-0.29}$\\ 
           NFW   &  $ 1.36^{+0.45}_{-0.29}$ & $ 30.2^{+12.1}_{-8.8}$     & $ 1.82^{+0.49}_{-0.39}$  &  $-0.22^{+0.18}_{-0.22}$\\ 
           HYB  &  $ 3.02^{+1.21}_{-0.78}$ & $ >117.5 $   & $ 1.08^{+0.31}_{-0.24}$  &  $-0.15^{+0.13}_{-0.25}$
\enddata
\end{deluxetable}

Here, we confine ourselves to investigating specific models for the halo density indices $(\alpha, \delta)$.  The model with $(\alpha, \delta)=(-2,0)$ corresponds to the singular isothermal sphere (hereafter SIS), and $(\alpha, \delta)=(-1,-1)$ is well known as the Navarro--Frenk--White profiles (hereafter NFW; Navarro et al. 1996, 1997), which can reproduce cosmological $N$-body simulations well. 
These two profiles are routinely used in galactic mass modeling and other studies (e.g., Evans \& Wilkinson 2000; Xue et al. 2008; Watkins et al. 2010).
Additionally, we examine another density profile of a galaxy-sized dark halo with $(\alpha, \delta)=(-2,-0.5)$.
This is a combination of SIS and NFW (hereafter hybrid profile, HYB), where its inner and outer parts are modeled by an isothermal density profile with $\rho\propto r^{-2}$ and by an NFW profile with $\rho\propto r^{-3}$ respectively. This is motivated by the possibility that a large amount of baryons infalling into the inner parts of a dark halo may induce a deeper and steeper density profile there than an NFW profile with $\rho\propto r^{-1}$, while the outer parts remain unchanged.
For one of the halo parameters, $\rho_0$, we replace it with an enclosed mass within a radius of 200~kpc in a spherical limit
$\rho(r)$ with $m\rightarrow r$ for $Q\rightarrow1$, 
\begin{eqnarray}
M_{\rm(\leq200\hspace{1mm}kpc)} = 4\pi\int^{200}_{0}\rho(r)r^2dr. 
\end{eqnarray}
We restrict ourselves to the inner regions of the satellite distributions, $r \leq 200$ kpc; beyond this range, the stellar population is likely to be seriously incomplete. 

In this work, in order to be determined by fitting to the observed line-of-sight velocity dispersion,
we adopt four parameters ($Q, b_{\rm halo}, M_{\rm(\leq200\hspace{1mm}kpc)}$, $\beta_z$) for the NFW and HYB halo models and three parameters ($Q, b_{\rm halo}$ or $M_{\rm(\leq200\hspace{1mm}kpc)}$, $\beta_z$) for the SIS model, respectively.

\subsection{Maximum Likelihood Analysis}
To obtain these halo parameters of our mass models by comparing with observational data, we employ a maximum likelihood analysis and construct the probability distribution for obtaining a set of line-of-sight velocities of GCs and dSphs in M31, based on the method explained below.

The observed velocity dispersion at a given position is the sum of two components; the intrinsic dispersion from the velocity distribution function and random error from the uncertainty in the measurement of the velocity. 
The intrinsic velocity dispersion is determined by the set of parameters that mainly govern the dark halo model. 
The probability for obtaining a set of line-of-sight velocities, $v_{{\rm los},i}$ $(i=1...N)$, given the three theoretical parameters, is $P(v_{{\rm los},i}|Q,b_{\rm halo},M_{\rm(\leq200\hspace{1mm}kpc)}, \beta_z)$. Supposing that the form of velocity distribution function is Gaussian, the probability distribution can be derived by
 \begin{eqnarray}
 \begin{split}
 P(v_{{\rm los},i}|Q,b_{\rm halo},&M_{\rm(\leq200\hspace{1mm}kpc)}, \beta_z) =  \\ 
 & \prod^{N}_{i=1}\frac{1}{\sqrt{2\pi\sigma^2_{i}}}\exp\Biggl[-\frac{1}{2}\frac{(v_{{\rm los},i}-u)^2}{\sigma^2_i}\Biggr],
  \end{split}
 \end{eqnarray}
where $v_{{\rm los},i}$ is a line-of-sight velocity in GSR and $u$ is defined as the systemic velocity of the Andromeda galaxy, $u=-123$ km~s$^{-1}$ (Rubin \& D'Odorico 1969). $N$ is the number of objects in the galaxy with line-of-sight velocity measurements, and $\sigma^2_i$ is the total variance at the projected position of the $i{\rm th}$ object given by $\sigma^2_i=\sigma^2_{t,i}+\sigma^2_{m,i}$, where $\sigma^2_{m,i}$ is the variance from the measurement uncertainty and $\sigma^2_{t,i}$ is the theoretical velocity dispersion derived from Jeans equations.
For this expression, we assume no correlations between the theoretical and measured dispersions. 
From the Bayes theorem, $P(v_{{\rm los},i}|Q,b_{\rm halo},M_{\rm(\leq200\hspace{1mm}kpc)}, \beta_z)$ in Equation (8) is proportional to the probability of the parameters given the data, $P(v_{{\rm los},i}|Q,b_{\rm halo},M_{\rm(\leq200\hspace{1mm}kpc)}, \beta_z) \propto P(Q,b_{\rm halo},M_{\rm(\leq200\hspace{1mm}kpc)}, \beta_z |v_{{\rm los},i})$. When considered as a function of the parameters, it can be defined as the likelihood function for the parameters.  
 \begin{figure}[!!t]
\begin{center}
  \includegraphics[scale=0.7]{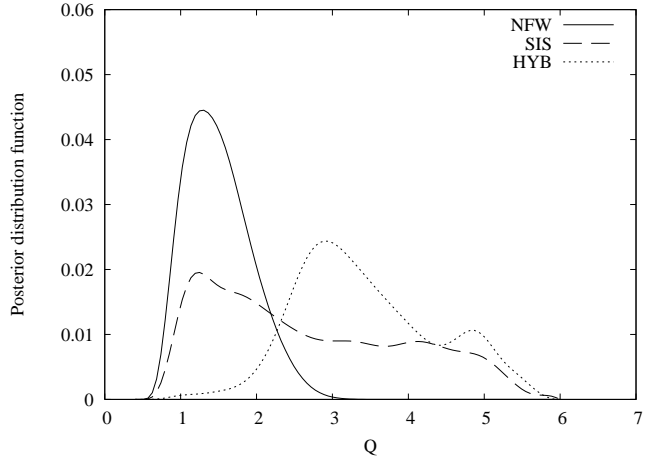}
 \end{center}
 \caption{Marginalized posterior distributions of the axial ratio, $Q$, of the dark halo. The SIS, NFW, and HYB models are indicated by solid, dashed, and dotted lines, respectively.}
 \label{fig:fig3}
\end{figure}

To do this analysis, we utilize Markov Chain Monte Carlo (MCMC) techniques with the standard Metropolis--Hasting algorithm (Metropolis et al. 1953; Hastings 1970). In a practical MCMC method, we calculate the likelihood $P({\bf M}|v_{{\rm los},i})$ for the current set of model parameters of ${\bf M} = (Q,b_{\rm halo},M_{\rm(\leq200\hspace{1mm}kpc)}, \beta_z )$. Then the next set of $\bf{M}^{\prime}$ is calculated by adding small random fluctuations to the previous $\bf{M}$, and the likelihood of $P^{\prime}({\bf M}^{\prime}|v_{{\rm los},i})$ is calculated.
If $P^{\prime}({\bf M}^{\prime}|v_{{\rm los},i})/P({\bf M}|v_{{\rm los},i})\geq1$, then the next set of model parameters $\bf{M}^{\prime}$ is accepted. 
If not, we draw a random variable $U$, which has a uniform probability between zero to one, and we accept $\bf{M}^{\prime}$ in the case of $P^{\prime}({\bf M}^{\prime}|v_{{\rm los},i})/ P({\bf M}|v_{{\rm los},i}) > U$. In other cases (i.e., $P^{\prime}({\bf M}^{\prime}|v_{{\rm los},i})/ P({\bf M}|v_{{\rm los},i}) \leq U$), $\bf{M}^{\prime}$ is rejected and the parameter set remains as the previous one $\bf{M}$.
These procedures are iterated for a large number of trials (at least $\sim 10^5$). The best estimate of a parameter set and their errors are obtained by taking means and standard deviations of the sample of $\bf{M}$ in the  MCMC trials, after removing early trials in the initial ``burn-in'' phase.

\begin{figure*}[t!!]
\begin{tabular}{cccc}
 \begin{minipage}{0.3\hsize}
  \begin{center}
   \includegraphics[width=57mm]{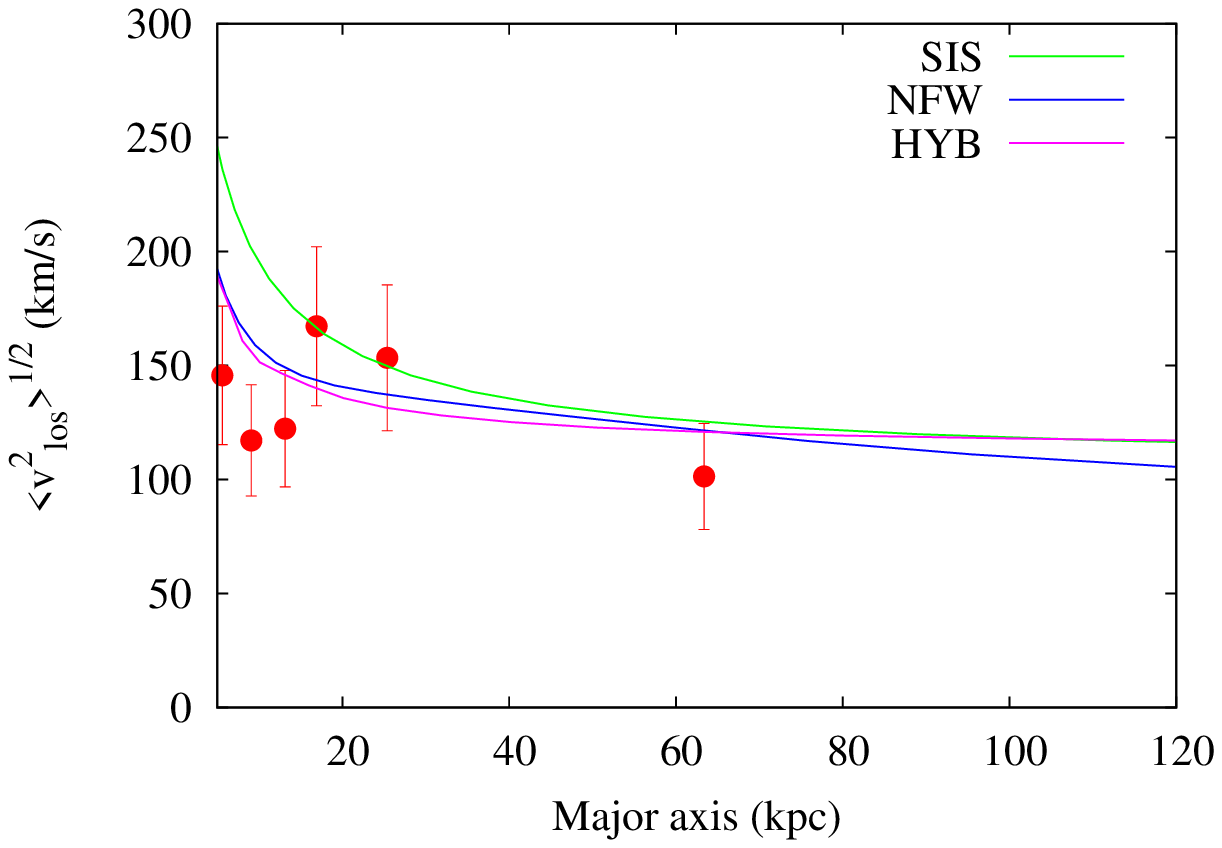}
  \end{center}
  \label{fig:one}
 \end{minipage}
  \begin{minipage}{0.3\hsize}
  \begin{center}
   \includegraphics[width=57mm]{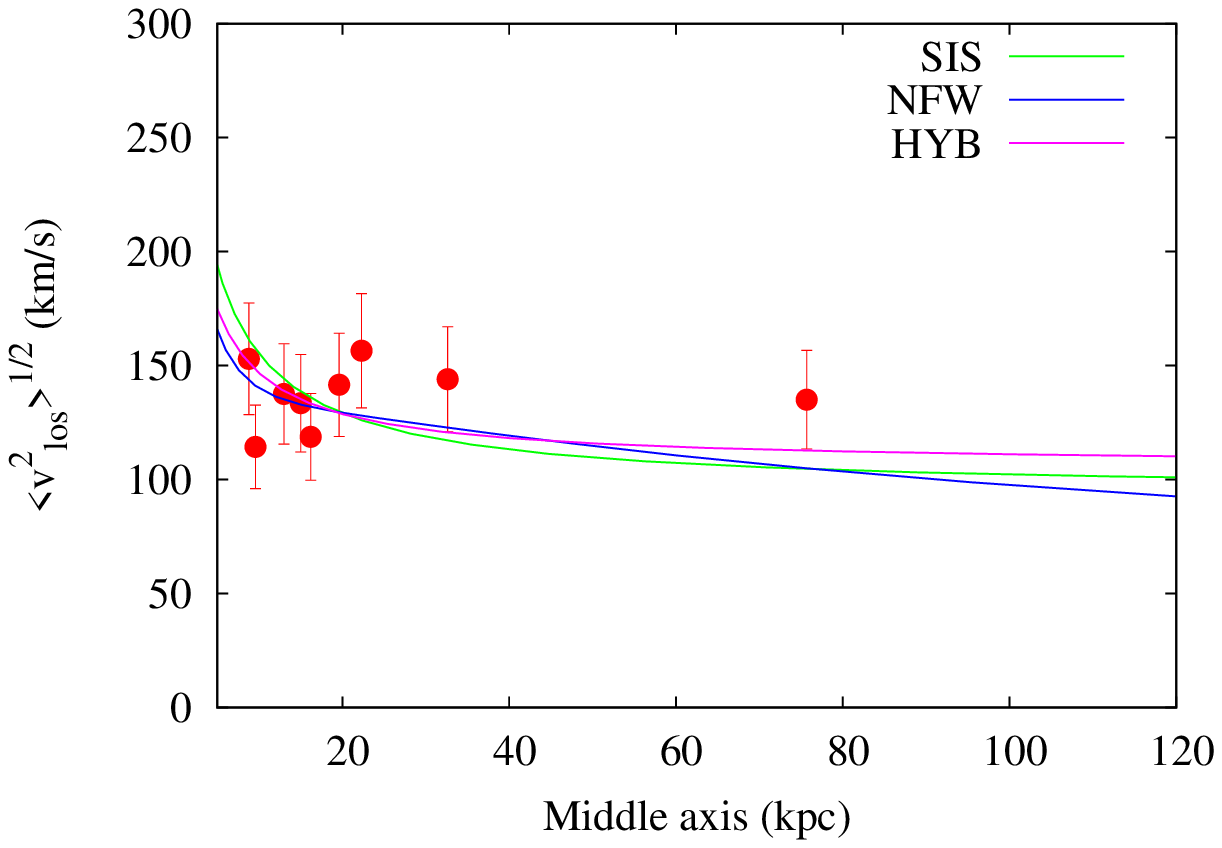}
  \end{center}
  \label{fig:two}
 \end{minipage}
  \begin{minipage}{0.3\hsize}
  \begin{center}
   \includegraphics[width=57mm]{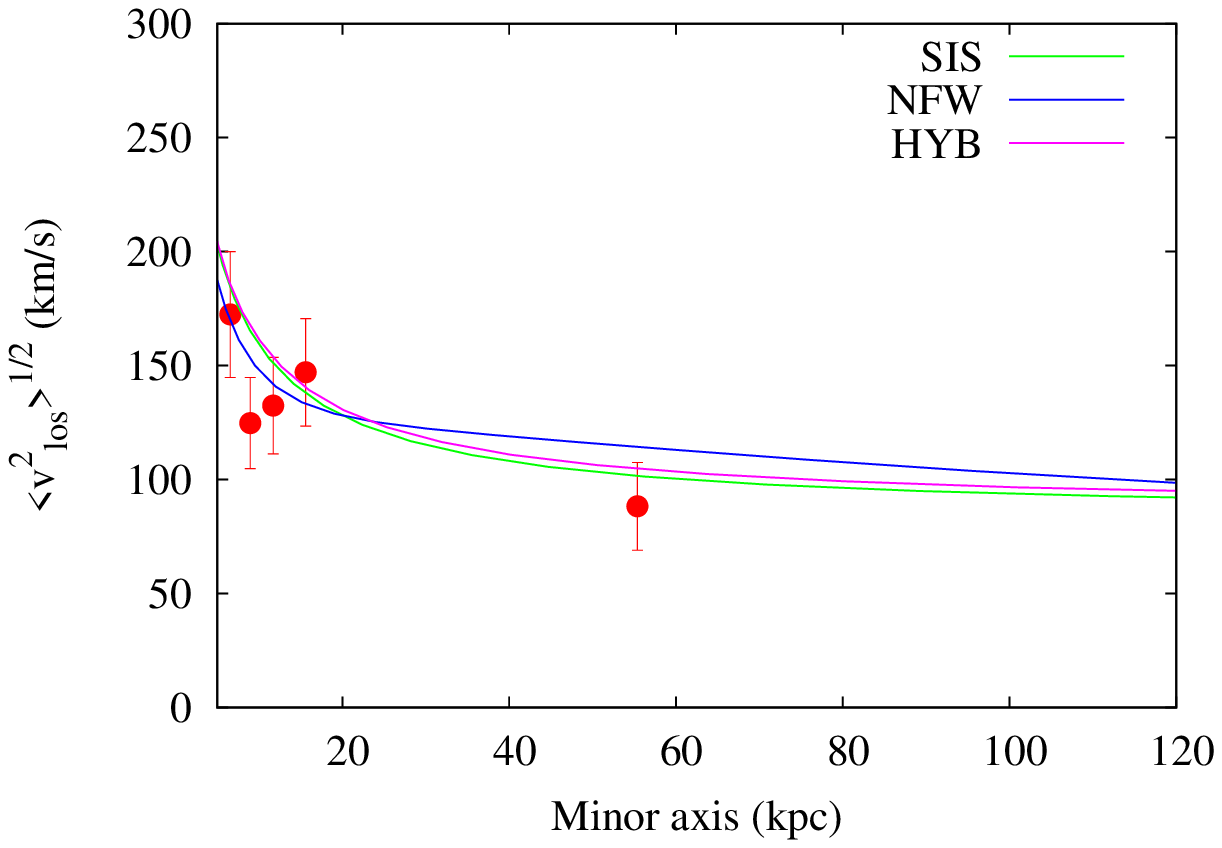}
  \end{center}
  \label{fig:three}
 \end{minipage}
 \end{tabular}
 \caption{Best-fit profiles of line-of-sight velocity dispersion along the major (left), middle (center), and minor (right) axes for the M31 halo. Red marks with error bars denote observed line-of-sight velocity dispersions along the major, middle, and minor axes, respectively. SIS, NFW, and HYB models are indicated by green, blue, and purple solid lines, respectively.}
\label{fig:fig4}
\end{figure*}

\section{RESULTS} 
We examine the allowed values for the entire parameter space for the dark matter halo to obtain the best fits to the spatial and kinematical data of the halo populations in M31. 
First, we set limits on the global shape of the dark halo characterized by an axial ratio, $Q$. 
Table 3 tabulates the best-fit results for the dark halo parameters that we obtain from the maximum likelihood method. Figure 3 shows the posterior distribution function as a function of $Q$ for the cases of the SIS, NFW, and HYB models, respectively. 
Figure~4 shows the most likely profiles of line-of-sight velocity dispersions along the
major, minor, and middle (which is defined at 45$^{\circ}$ from the major axis) axes for SIS (green line), NFW (blue line), and HYB (purple line), respectively. We adopt the best-fit parameters, which are obtained from the above maximum likelihood method, to draw these profiles.
It is clear from Column 3 in Table 3 and Figures 3 and 4 that the best-fit shapes of the dark matter halo in M31 are generally {\it prolate} with $Q>1$ for all of dark halo density models. 
The reason for these results is as follows.
Our numerical experiments for the solution of Jeans equations suggest that for a steep three-dimensional density profile of halo tracers, $\nu \propto m^{-4}_{\ast}$, where $m_{\ast}^2=R^2+z^2/q^2$ with $q$ larger than unity, their $\sigma_{\rm los}$ profile is such that it is sharply peaked near
the M31 center, then decreasing rapidly with increasing projected distance along the minor axis;
such a rapidly decreasing $\sigma_{\rm los}$ profile is needed to be compatible with
the effect of a steep $\nu$ profile in Jeans equations.  On the other hand, $\sigma_{\rm los}$ obtained from our sample in M31 shows a rather flat profile. To reproduce such an observed flat $\sigma_{\rm los}$ profile, a background dark halo is required to have a prolate shape, $Q > 1$, so that the spatial gradient of gravitational force along the minor axis remains small.
Otherwise, if $Q < 1$, the spatial gradient of $\sigma_{\rm los}$ profile is made much steeper, i.e., deviating more from the observed profile.

We investigate the degeneracy in model fitting for determining the four parameters $(Q,b_{\rm halo},M_{\leq {\rm 200kpc}}, \beta_z)$. In Figures 5--7, we present 68 \%, 95 \%, and 99 \% confidence levels of contours in the two-dimensional plane of $Q$--$b_{\rm halo}$, $Q$--$M_{\leq {\rm 200kpc}}$, $Q$--$\beta_z$ and $b_{\rm halo}$--$M_{\leq {\rm 200kpc}}$ for the NFW, SIS, and HYB models, respectively. 
In the NFW model, as shown in Figure 5, it follows that $Q$--$b_{\rm halo}$, $Q$--$M_{\leq {\rm 200kpc}}$ and $Q$--$\beta_z$ contour maps show only a weak degeneracy with respect to $Q$, with a ridge at larger $Q$ values, so that a prolate dark halo is most probable without strong degeneracies.
Contrary to this, as seen from the $b_{\rm halo}$--$M_{\leq {\rm 200kpc}}$ map, there exists an obvious degeneracy between these two parameters; both of these parameters affect the total amplitude of velocity dispersions. Thus, with available data alone it is difficult to break this $b_{\rm halo}$--$M_{\leq {\rm 200kpc}}$ degeneracy.
On the other hand, it is more difficult to determine the best-fit value of $Q$ in the SIS and HYB models because of a widely distributed confidence interval as shown in Figures 6 and 7. As for the HYB model, although we are only able to constrain a  lower limit on $b_{\rm halo}$, this has only a small impact on the other two parameters.
\begin{deluxetable}{lccc}[!T]
\label{tb:tab4}
\tablecolumns{4}
\tablewidth{3.0in}
\tablecaption{Best-fit $Q$ Values for Different Halo Selections}
\tablehead{ Data & NFW  & SIS  & HYB}
\startdata
  DATA~1  & $ Q=1.32^{+0.55}_{-0.24}$ & $ Q=1.55^{+0.44}_{-0.25}$ &  $ Q=2.99^{+0.98}_{-0.66}$\\ 
  DATA~2  & $ Q=1.24^{+0.30}_{-0.33}$ & $ Q=1.43^{+0.25}_{-0.31}$ &  $ Q=2.70^{+1.12}_{-0.73}$\\
  DATA~3  & $ Q=1.92^{+1.33}_{-0.90}$ & $ Q=2.69^{+1.44}_{-1.21}$ &  $ Q=3.34^{+1.60}_{-1.30}$\\
  Full data & $ Q=1.36^{+0.45}_{-0.29}$ &  $ Q=1.60^{+0.38}_{-0.51}$ & $ Q=3.02^{+1.21}_{-0.78}$
\enddata
\end{deluxetable}

We compare our mass estimate of Andromeda with previous work. 
Evans \& Wilkinson (2000) estimated that the total mass of M31 is $1.23^{+1.8}_{-0.6}\times10^{12}M_{\odot}$ with a scale length of $\sim 90$~kpc derived from positions and radial velocities of its satellite galaxies, distant GCs, and planetary nebulae.
Watkins et al. (2010) found that the mass of M31 within the inner 200~kpc is $1.1\sim1.5\times10^{12}M_{\odot}$, including the uncertainty of velocity anisotropy. Recently, Veljanoski et al. (2013) presented their mass estimation based on a kinematic analysis of far outer halo GCs in M31 and found the mass within the inner 200~kpc of $(1.2$-$1.5)\pm0.2\times10^{12}M_{\odot}$.
As shown in Table 3, $M_{\leq {\rm 200kpc}}$ obtained here is almost the same as or slightly more massive than previous mass estimates. 

As indicated above, we find the prolate dark halo of M31 most likely, regardless of assumed density profiles.
However, it is unclear to what extent other observational parameters (i.e., tracer distribution parameters, $\gamma$, and $q^{\prime}$) as well as sample selections affect the covariance between halo parameters.
Therefore, in what follows, we evaluate dark halo parameters taking into account all the above possible factors and show that M31's dark halo is prolate even considering these effects on the best-fit $Q$ value. 

In order to inspect the impact of data sampling on fitting results, we perform the same velocity analysis for the three different samples: DATA~1 contains halo tracers within 150~kpc from the center of M31 and DATA~2 and DATA~3 contain only GC and dSph tracers, respectively.
As shown in Table 4, we find that the best-fit $Q$ is always larger than unity independent of these sample selections, although the values of the best-fit $Q$ are varied depending on the assumed models.

We repeat the maximum likelihood analysis to obtain the best-fit parameters, including the parameters for a tracer distribution, because the density profile of halo tracers, $\nu$, also affects the best-fit $Q$ value of the dark halo through Jeans equations.
To investigate this issue, we include tracer distribution parameters, $\gamma$ and $q^{\prime}$, in the likelihood function, i.e., we replace  $P(v_{{\rm los},i}|Q,b_{\rm halo},M_{\rm(\leq200\hspace{1mm}kpc)}, \beta_z)$ with  $P(v_{{\rm los},i}|Q,b_{\rm halo},M_{\rm(\leq200\hspace{1mm}kpc)}, \beta_z, \gamma, q^{\prime})$.
Taking the case of the NFW model as an example, we find that the best-fit values of $Q$, $\gamma$, and $q^{\prime}$ are $Q=1.50\pm0.39$, $\gamma=-3.27\pm0.38$, and $q^{\prime}=1.09\pm0.15$, respectively. 
We also obtain similar results for other models (SIS: $Q=2.70\pm1.30$, HYB: $Q=3.47\pm0.92$). 
Consequently, we find that the best-fit axial ratios of M31's dark halo, $Q$, are always larger than unity even considering the impacts of tracer distribution parameters in our maximum likelihood analysis. It is also found that the best-fit values of $\gamma$ and $q^{\prime}$ are roughly in agreement with those we have obtained separately for each, as shown in Section~2.3.

\section{DISCUSSION AND CONCLUDING REMARKS}
We have shown that our best-fit models for M31 reveal a prolate dark halo under the assumption of axisymmetry.
This result also indicates that the major axis of the M31 dark halo is perpendicular to its disk and nearly aligned with  the planer distribution of a subgroup of its satellites (Ibata et al. 2013).
These results are consistent with the CDM prediction that the satellites are distributed anisotropically and preferentially located along major axes of their galactic host halos, and that the host halos are elongated along the dominant filaments of the cosmic web (e.g., Kang et al. 2005; Zentner et al. 2005; Libeskind et al. 2005).

However, these $N$-body simulations based on a standard CDM theory of collisionless systems predict that dark halos are mildly flattened shapes: intermediate-to-major axis ratios, $b/a\sim0.7$, and minor-to-major axis ratios, $c/a\sim0.5$ (supposing $a\geq b\geq c$). Moreover, Kazantzidis et al. (2010, and reference therein) suggested that when infalling cooled baryonic is taken into account, dark halos tend to be more spherical than purely collisionless cases. 
Therefore, the shape of the M31 dark halo obtained from observations is more elongated than theoretical predictions for MW-sized, CDM dark halos.
\begin{figure*}[t!!]
 \begin{minipage}{0.5\hsize}
  \begin{center}
   \includegraphics[width=80mm]{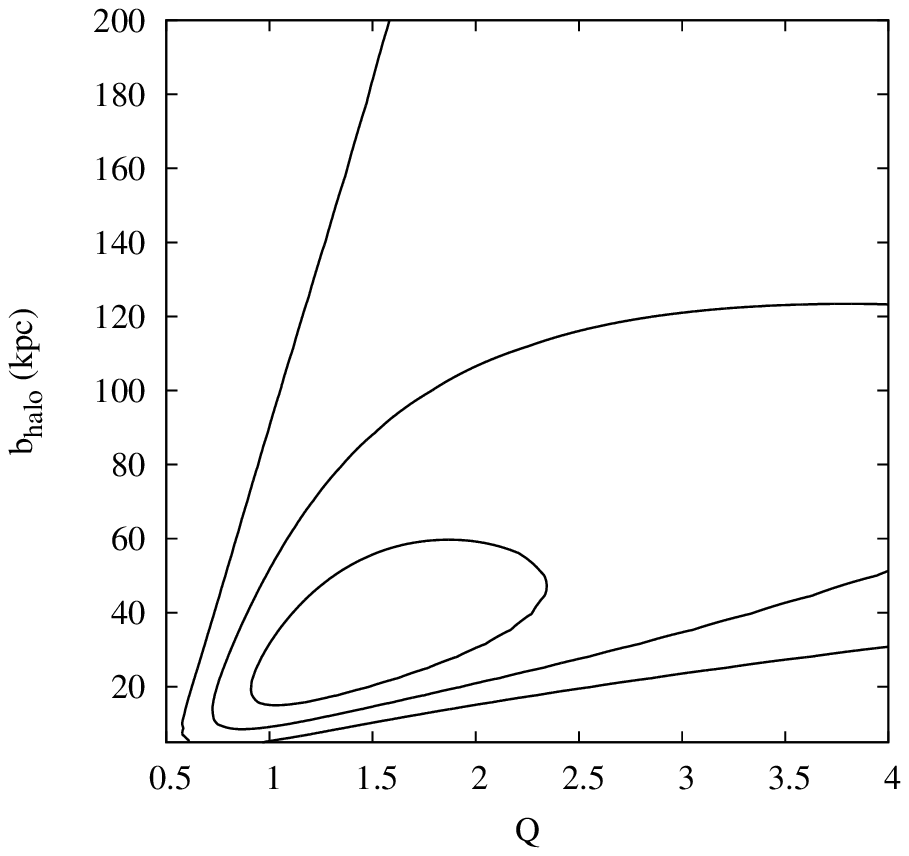}
  \end{center}
 \end{minipage}
 \begin{minipage}{0.5\hsize}
  \begin{center}
   \includegraphics[width=80mm]{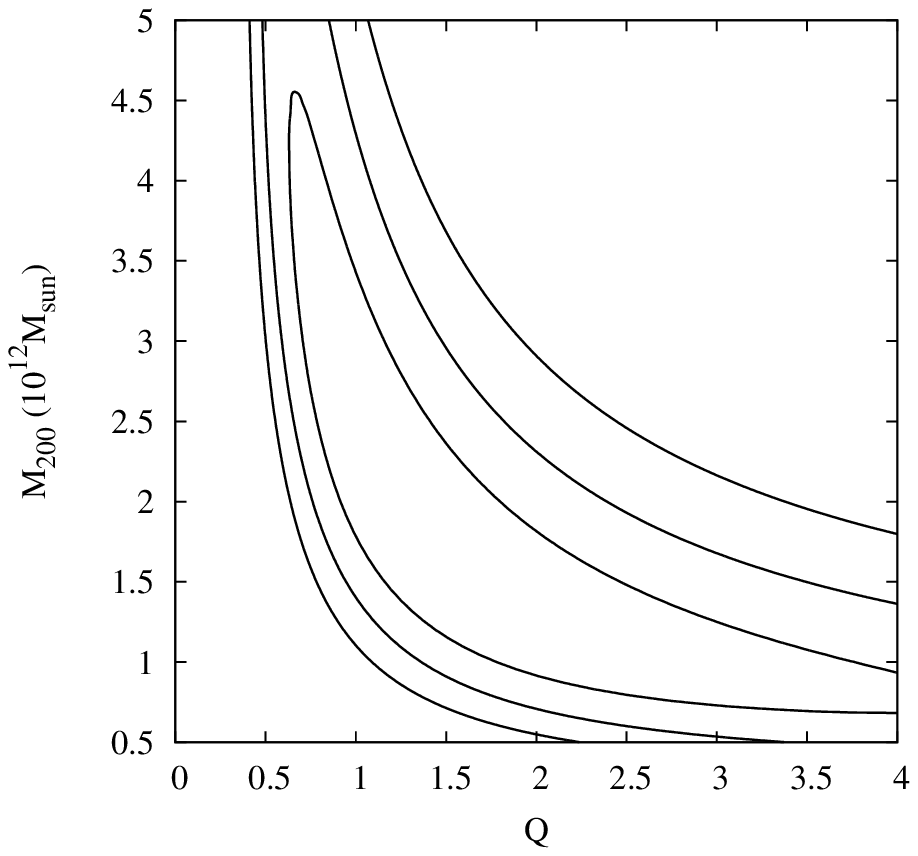}
  \end{center}
 \end{minipage}
  \begin{minipage}{0.5\hsize}
  \begin{center}
   \includegraphics[width=80mm]{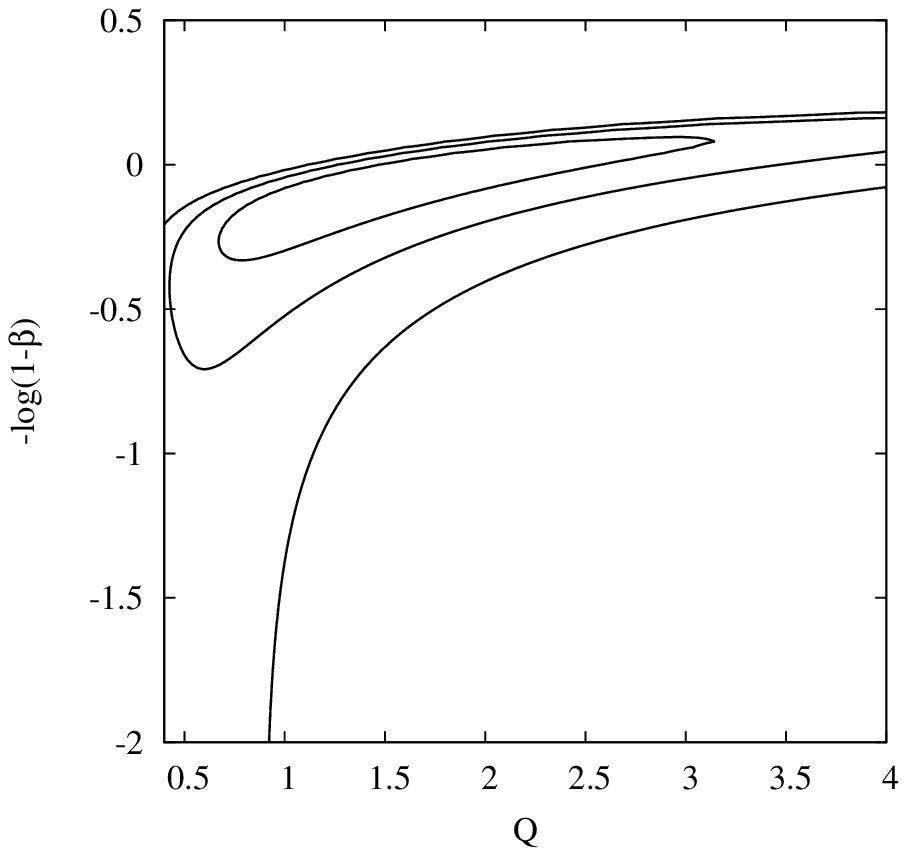}
  \end{center}
 \end{minipage}
 \begin{minipage}{0.5\hsize}
  \begin{center}
   \includegraphics[width=80mm]{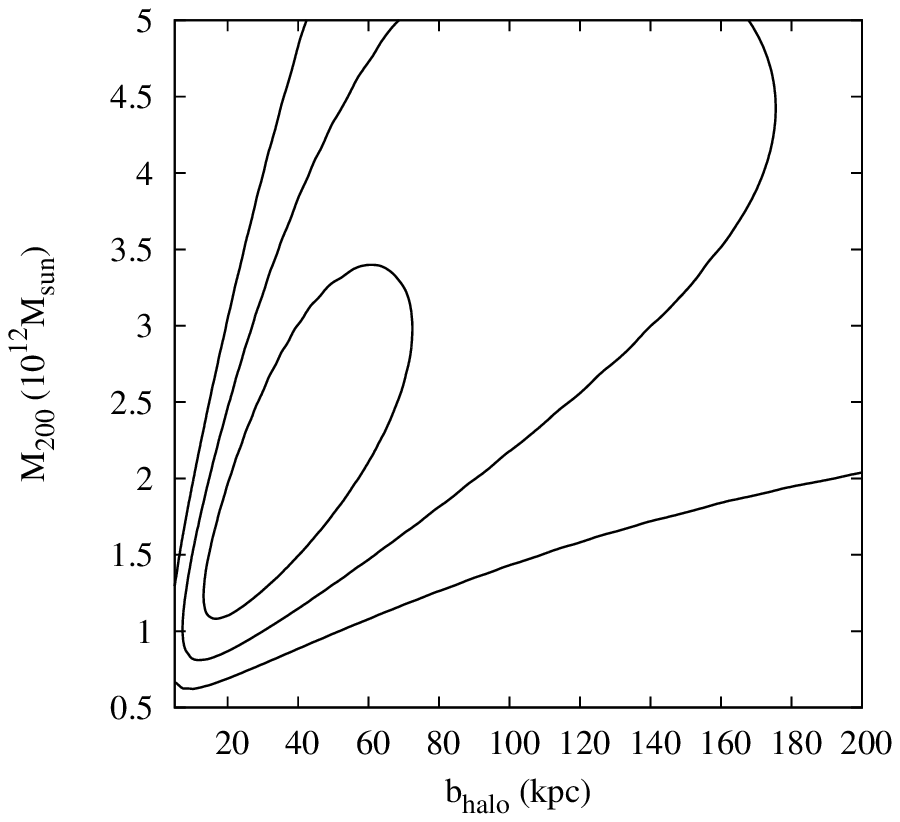}
  \end{center}
 \end{minipage}
 \caption{Likelihood contours for dark halo parameters based on our NFW model. Contours show 68\%, 95\%, and 99\% confidence levels. Clockwise from top left:  $Q$--$b_{\rm halo}$, $Q$--$M_{\leq 200{\rm kpc}}$, $Q$--$\beta_z$, and $b_{\rm halo}$--$M_{\leq 200{\rm kpc}}$.}
  \label{fig:fig5}
\end{figure*}
\begin{figure*}[htbp]
 \begin{minipage}{0.5\hsize}
  \begin{center}
   \includegraphics[width=80mm]{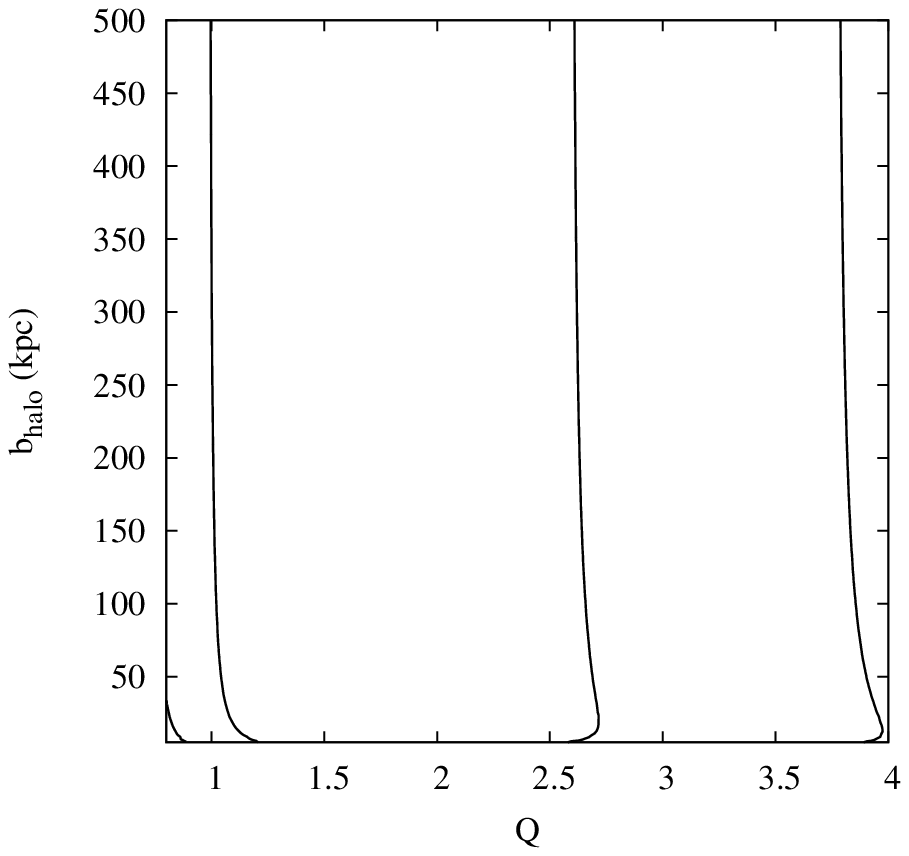}
  \end{center}
 \end{minipage}
 \begin{minipage}{0.5\hsize}
  \begin{center}
   \includegraphics[width=80mm]{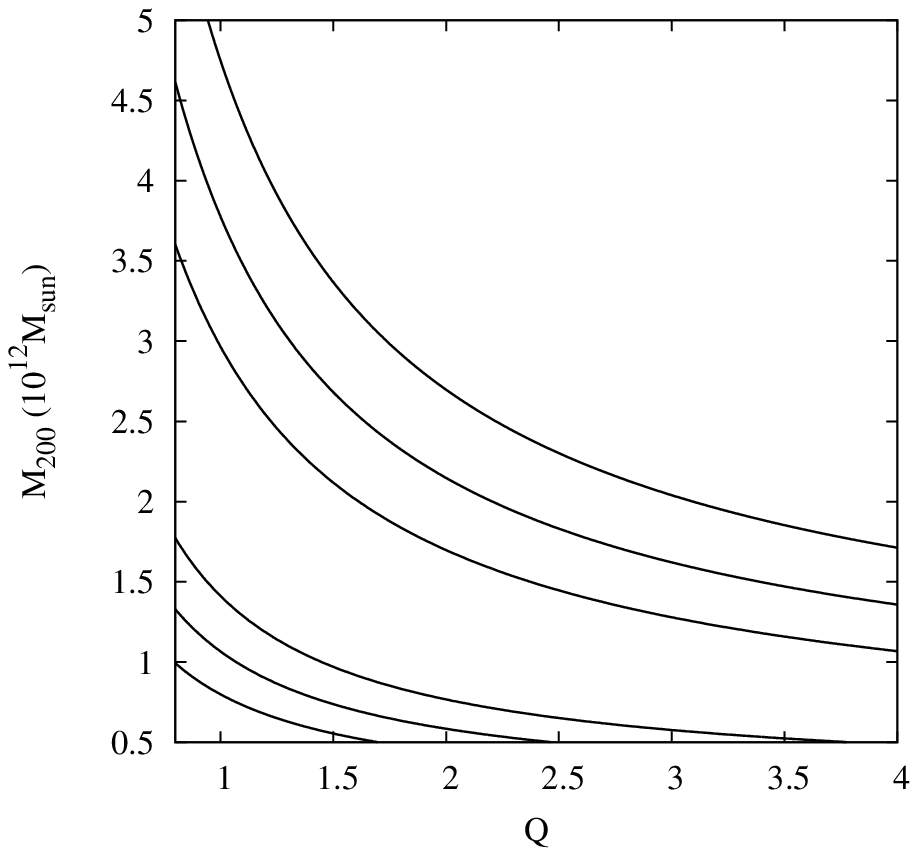}
  \end{center}
 \end{minipage}
  \begin{minipage}{0.5\hsize}
  \begin{center}
   \includegraphics[width=80mm]{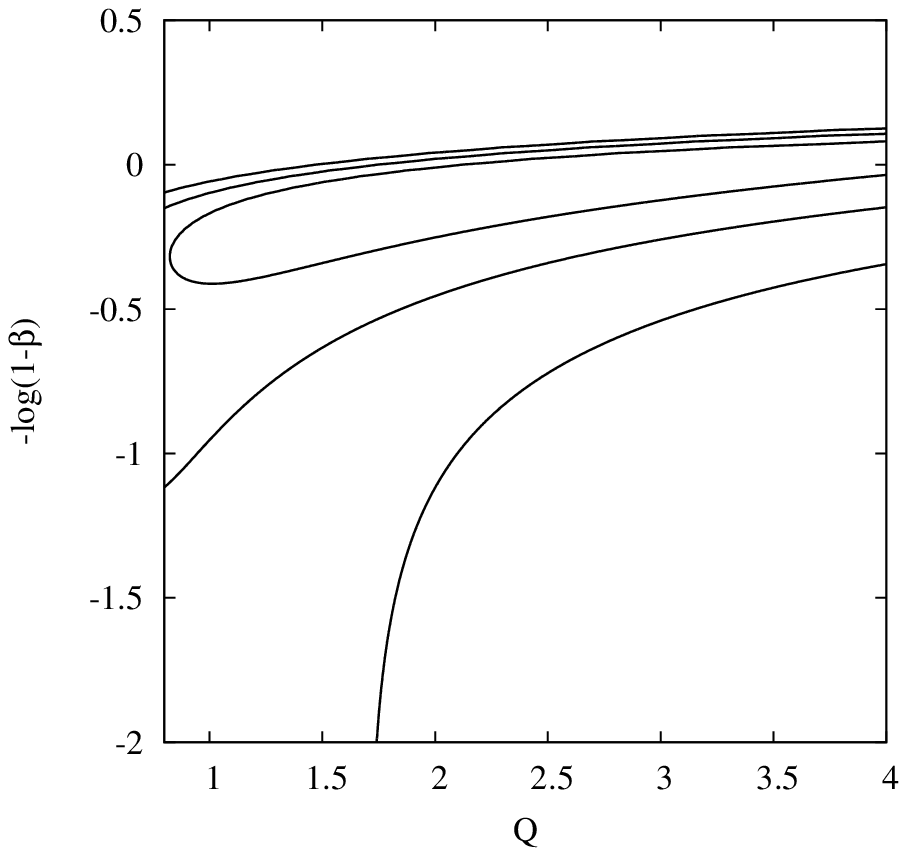}
  \end{center}
 \end{minipage}
 \caption{Same as Figure 5, but for the case of the SIS model.}
  \label{fig:fig6}
\end{figure*}
\begin{figure*}[htbp]
 \begin{minipage}{0.5\hsize}
  \begin{center}
   \includegraphics[width=80mm]{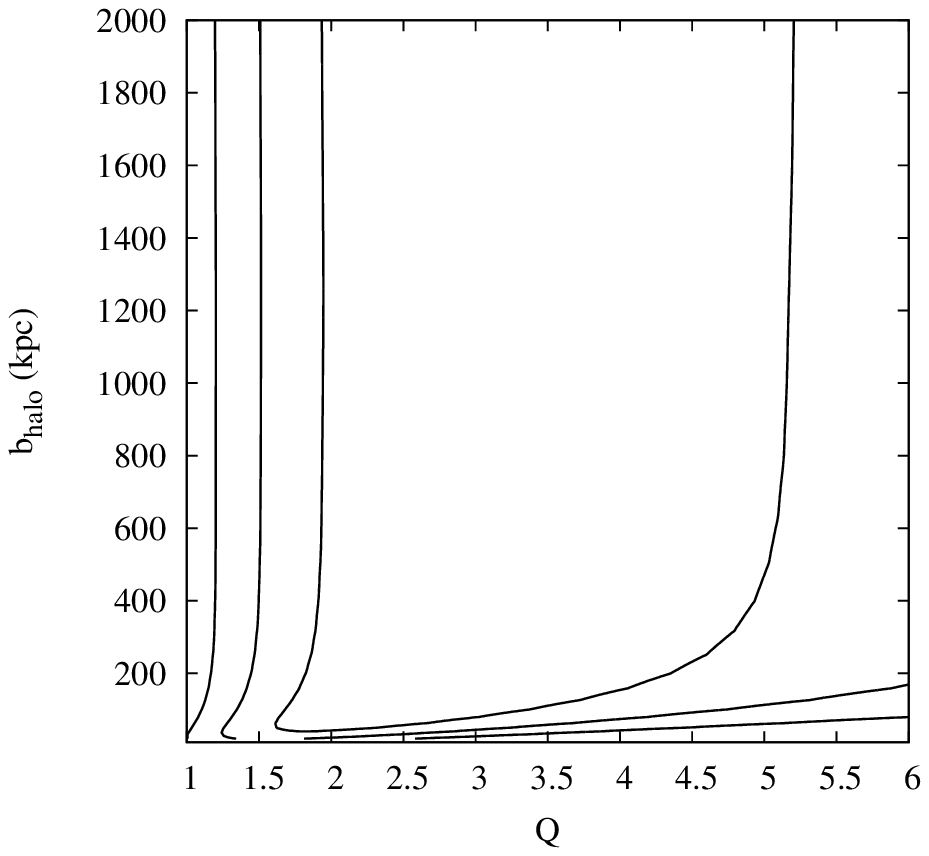}
  \end{center}
 \end{minipage}
 \begin{minipage}{0.5\hsize}
  \begin{center}
   \includegraphics[width=80mm]{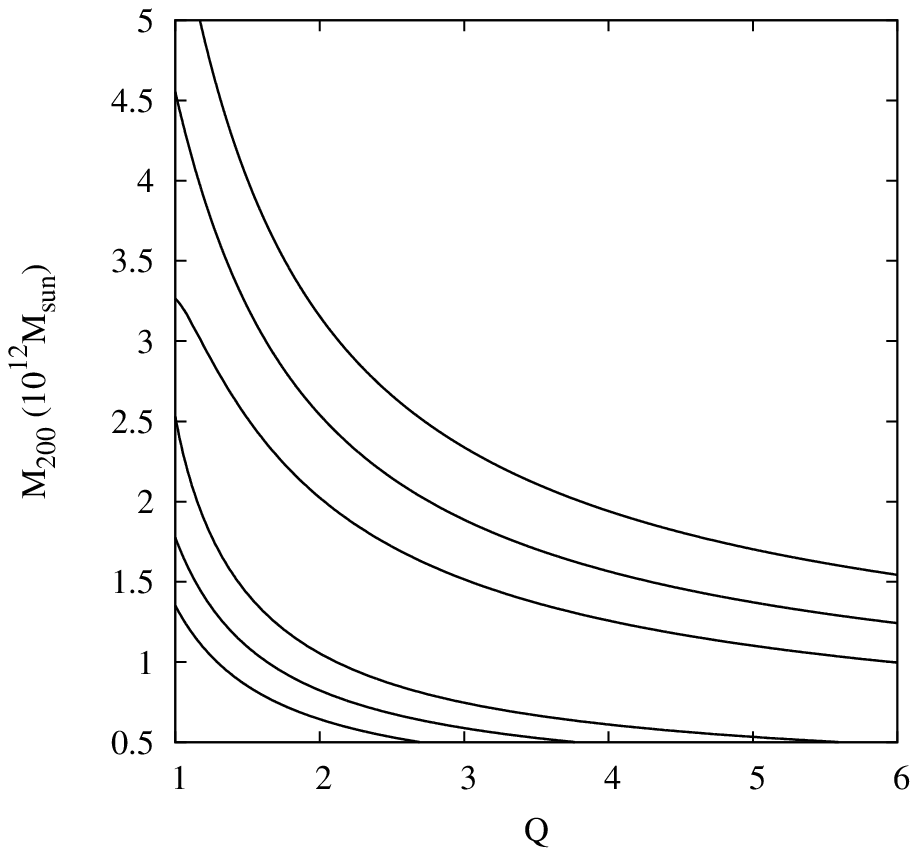}
  \end{center}
 \end{minipage}
  \begin{minipage}{0.5\hsize}
  \begin{center}
   \includegraphics[width=80mm]{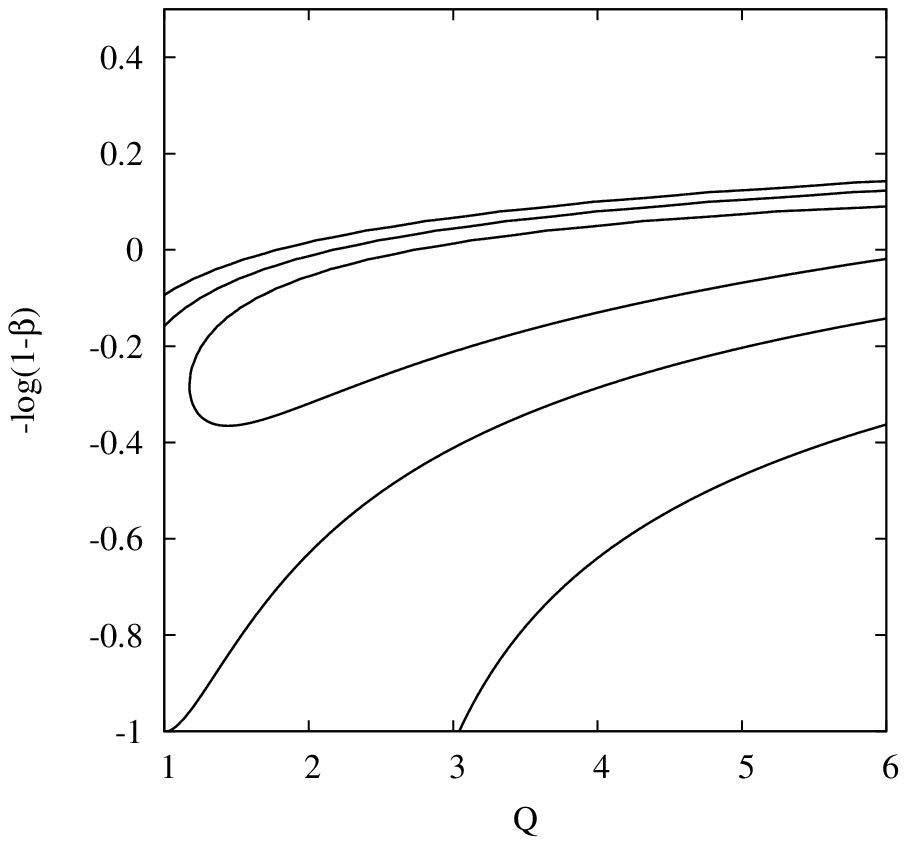}
  \end{center}
 \end{minipage}
 \begin{minipage}{0.5\hsize}
  \begin{center}
   \includegraphics[width=80mm]{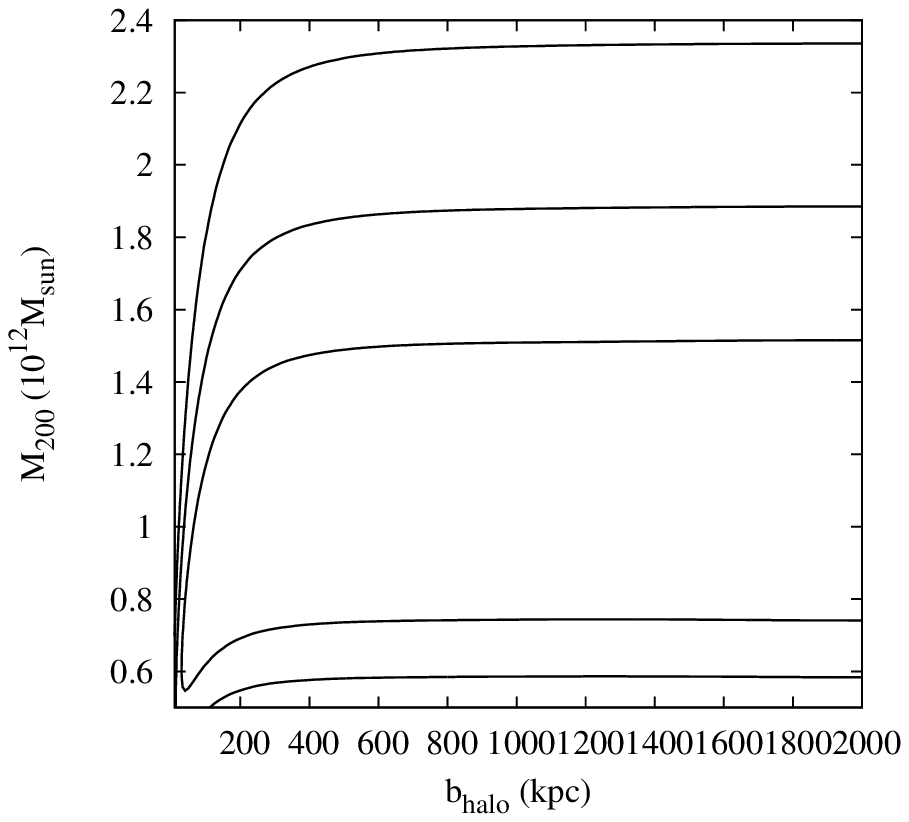}
  \end{center}
\end{minipage}
 \caption{Same as Figure 5, but for the case of the HYB model.}
  \label{fig:fig7}
\end{figure*}

In comparison, for the MW, the spatial and kinematical distributions of the Sagittarius tidal stream provide useful limits on the halo shape. Recent analysis of the Sagittarius stream shows that the MW halo is triaxial or a nearly oblate ellipsoid whose minor axis is contained within the Galactic disk plane (e.g., Johnston et al. 2005; Law et al. 2009; Law \& Majewski 2010). However, some issues remain, including the dependency of initial mass of the Sagittarius dwarf galaxy on the result in Helmi (2004), who found that MW dark halo is either oblate or prolate. 
In the work of Law \& Majewski (2010), the nature of  the bifurcation of the Sagittarius stream (Koposov et al. 2013) remains unanswered, which may affect the orbital analysis of the Sagittarius dwarf galaxy.

Dynamical effects of dark matter subhalos and/or luminous satellites accretion play an important role in describing formation history of a galaxy, including the formation of a thick disk (e.g., Hayashi \& Chiba 2006; Villalobos \& Helmi 2008; Kazantzidis et al. 2009; Villalobos et al. 2010) and of a stellar halo (e.g., Helmi \& White 1999; Bullock \& Johnston 2005; Font et al. 2011). 
We note that many of the previous mass models assumed a spherical host halo, even though the non-spherical shape of a dark halo and the resulting orbits of satellites within it may have an important consequence on the evolution of luminous stellar components.
Indeed, using a cosmological hydrodynamical simulation, Valluri et al. (2013) found that about a half of stellar orbits are short axis tube orbits and the rest belong to orbits of characterized triaxial potential, and also the type of orbits are different between young and old tidal debris.  
This may modify both the heating of the stellar disk and the efficiency of tidal disruption of satellites, thus the formation process of the stellar halo. 
Also, Carlberg~et~al. (2011) investigated the M31 stellar stream extending about 120~kpc northwest from the center of M31 (Richardson~et~al. 2011) to constrain the abundance of dark matter subhalos in it, which may induce the gaps in the stream through dynamical effects.
Their analysis is based on a spherical host halo in M31, but the use of a prolate halo suggested from the current work may affect the dynamical influence of subhalos on the stream, through the different orbital realizations of both subhalos and the stream.   
Thus, our result is profound in understanding the dynamics of halo tracers and the orbital evolutions of tidal stellar streams, which play important roles in extracting the abundance of CDM subhalos through their dynamical effects on stream structures.  

In summary, using and generalizing the axisymmetric mass models developed by HC12, we set new limits on the global shape and density profile of the Andromeda halo and compare our results with theoretical prediction of CDM models. This is motivated by the fact that CDM models predict non-spherical virialized dark halos, which reflect the process of mass assembly in the galactic scale.
Applying our models to the latest kinematic data of GCs and dSphs in the Andromeda halo, we have found that the most plausible cases for Andromeda yield a prolate shape for its dark halo.
Furthermore, the prolate dark halo is consistent with theoretical predictions in which the satellites are distributed anisotropically and preferentially located along major axes of their galactic host halos.
In the near future, planned surveys of Andromeda's halo using Hyper Suprime-Cam and Prime Focus Spectrograph attached to the Subaru Telescope (Takada et al. 2014) will enable us to discover new halo objects (GCs, dSphs, and tidal streams) and measure their accurate kinematic data, thereby allowing us to obtain tighter limits on the dark halo distribution in Andromeda.
\acknowledgments
The authors thank the referee for constructive comments that have helped us to improve our paper.
This work has been supported in part by a Grant-in-Aid for Scientific Research (20340039, 18072001) of the Ministry
of Education, Culture, Sports, Science and Technology in Japan and by the JSPS Core-to-Core Program ``International Research Network for Dark Energy''.
\clearpage


\end{document}